\documentclass[12pt]{iopart}
\usepackage{graphics}
\usepackage{graphicx}
\usepackage{amsfonts}
\usepackage{amssymb}
\usepackage{cite}
\usepackage{xcolor}
\usepackage{dsfont}
\usepackage[latin1]{inputenc}
\usepackage{amsthm}
\usepackage{amsmath}
\usepackage{amscd}
\usepackage{iopams}

\def\beqn{\begin{eqnarray}}
\def\eeqn{\end{eqnarray}}
\def\aap{a^\dagger}
\def\aam{a}
\def\hw{\hbar \omega}
\def\hw4{ \frac {\hbar \omega}{4}}

\def\uni{{\bf i}}
\def\re{{\rm {e}}}

\def\th{\theta}

\def\al{\alpha}
\def\be{\beta}
\def\sg{\sigma}
\def\Om{\Omega}
\def\Omt{|\Omega| t}
\def\om{\omega}
\def\b0{b_0}
\def\p0{p_0}
\def\s0{\sigma_0}
\def\x0{x_0}
\def\g0{\gamma_0}
\def\bi{\textbf{i}}
\def\wid{\widetilde}
\def\ovl{\overline}

\def\hx{\hat{x}}
\def\hp{\hat{p}}

\def\pt{{\mathcal{PT}}}
\def\nonn{\nonumber \\}

\def\pt{{\cal{PT}}}
\def\nonn{\nonumber \\}
\def\lth{{\rm e}^{\uni \left(\th-\frac \pi 4 \right)} \frac {|\sg|}{b_0}  x}
\def\llth{{\rm e}^{\uni \left(\th-\frac \pi 4 \right)} \frac {|\sg|}{b_0}  L}

\def\lpth{{\rm e}^{-\uni \left(\th-\frac \pi 4 \right)} \frac{b_0}{\hbar |\sg|} p}

\begin{document}

\title[Complex Scaling Method: the Swanson Hamiltonian in the broken PT-symmetry phase.]{Complex Scaling Method applied to the study of the Swanson Hamiltonian in the broken PT-symmetry phase.}

\author{ Viviano Fern\'andez$^{1}$, Romina Ram\'\i rez$^{2}$ , Marta Reboiro$^{3}$ }

\address{{\small\it $^{1}$} CONICET-Department of Physics, University of La Plata, Argentina}
\address{{\small\it $^{2}$} IAM, CONICET-CeMaLP, University of La Plata, Argentina}
\address{{\small\it $^{3}$}IFLP, CONICET-Department of Physics, University of La Plata, Argentina}

\ead{reboiro@fisica.unlp.edu.ar}

\begin{abstract}
In this work, we study the non-PT symmetry phase of the Swanson Hamiltonian in the framework of the Complex Scaling Method.
By constructing a bi-orthogonality relation, we apply the formalism of the response function to analyse the time evolution of different initial wave packages. The Wigner Functions and mean value of operators are evaluated as a function of time. We analyse in detail the time evolution in the neighbourhood of Exceptional Points. We derive a continuity equation for the system. We compare the results obtained using the Complex Scaling Method to the ones obtained by working in a Rigged Hilbert Space.
\end{abstract}


%
\submitto{\PS}

\maketitle

\section{Introduction}

In recent years, non-hermitian dynamics has become a field which attracted increasing interest \cite{ueda,meden,bagarellobook}. It has been successfully applied to a wide range of areas, such as quantum optics \cite{qopt0,qopt1,cv1,cv2}, condensed matter physics \cite{condense} and quantum field theory \cite{field0,field1}. In its beginnings, the motivation was focused on the extension of the quantum mechanics to include Parity-Time Reversal (\pt)-symmetry operators, which can display real spectrum \cite{bender0,bender1,bender6,bender2,bender3,bender4,bender5}. Specifically, these operators are a class of the so-called pseudo-hermitian operators \cite{AMostafazadeh,ali1}. They admit real spectrum or complex conjugate pairs of eigenvalues. 
For the real spectrum, the \pt-symmetry is exact, and the corresponding eigenfunctions obey \pt-symmetry. For complex conjugate pairs eigenvalues,  \pt-symmetry is broken, and the associated eigenfunctions are not \pt-symmetric. The border between both regions consists of Exceptional Points (EPs). At EPs, both the eigenvalues and the eigenfunctions are coalescent. From the pioneering work of C. Bender and co-workers, \cite{bender0}, the literature has grown exponentially, particularly in dealing with the exact \pt-symmetry phase and the new physics that arise in the neighbourhood of the EPs \cite{ueda,meden}. However, there are open questions describing the broken symmetry phase, such as some mathematical aspects of the formalism and the physical interpretation of results. 

The study of the spectrum in the broken symmetry phase involves eigenfunctions which are not square-integrable, they belong to the Rigged Hilbert Space \cite{gelfand,bhom}. 
An alternative to the construction of a Gelfand's Triplet is the implementation of the Complex Scaling Method (CMS) \cite{csm0,csm1,csm2,moiseyev0,moiseyevpr,moiseyev}. The CSM has been extensively used in dealing with scattering problems in different contexts, particularly in nuclear physics \cite{csm3,csm5}. This formalism consists of a dilatation transformation, which allows working with square-integrable functions for a given range of the parameter of dilatation. The price to pay for working with square integrable eigenfunction is that the transformed operators are not self-adjoint so a new inner product has to be introduced \cite{moiseyevpr,csm5}. The essentials of the formalism are given in the ABC theorem by Aguilar, Combes and Balslev \cite{csm0,csm1} and in the works of N. Moiseyev and coworkers \cite{moiseyevpr,csm5}.  

In this work, we study the non-PT symmetry phase of the Swanson Hamiltonian \cite{swanson,sw1,sw2,sw3,sw4,sw5,sw6,sw7,sw8,sw9,sw10,sw11,sw12,sw13,sw14,sw15,sw16,sw17,sw18,sw19,sw20,sw21,sw22,nosswrhs} in the framework of the Complex Scaling Method\cite{csm0,csm1,csm2,moiseyev0,moiseyevpr,moiseyev,noscsm}.  
By constructing a bi-orthogonality relation, we extend the formalism of the response function to analyse the time evolution of different initial wave packages under the action of a \pt-symmetric Hamiltonian \cite{response}. The Wigner Functions and mean value of operators are evaluated as a function of time. Particularly, we analyse time evolution in the vicinity of EPs. We derive a system continuity equation based on the proposed bi-orthogonal relation. 
We compare the results obtained using the Complex Scaling Method and the ones obtained by working in a Rigged Hilbert Space.

The work is organised as follows. In Section \ref{form}, we present the Swanson model and the methods we have adopted in its study. We introduce the Complex Scaling Method and in \ref{dis} and \ref{con} we derive the eigenfunctions and eigenvalues of the transformed Hamiltonian for the discrete and the continuum spectrum, respectively. In \ref{te}, the time evolution under the action of the transformed Hamiltonian is developed in the framework of the Response Function Formalism. In \ref{wig}, the Wigner Functions of the problem are presented and from them, the derivation of the density spatial distribution and the continuity equation is described. In \ref{results}, we depicted the results obtained for the time evolution of different wave packages. Conclusions are drawn in \ref{conclusions}. 

\section{Formalism}
\label{form}

To discuss the use of the CSM in the study of \pt-symmetry systems, we shall consider the Swanson Hamiltonian \cite{swanson}, which is given by

\begin{equation}
{\mathrm H}(\omega,\alpha,\beta)= \hbar \omega~ \left( \aap \aam+ \frac 12 \right) +
\hbar \alpha~ {\aam}^2+  \hbar \beta ~ {\aap}^2.
\label{hswsq}
\end{equation}

In Eq.(\ref{hswsq}) the energy scale is fixed by the real energy $\hbar \om$, where $\om$ can be interpreted as the frequency of a harmonic oscillator of mass $m_0$ and restoring parameter $k_0$, $\om=(k_0/m_0)^{1/2}$.
The quantities $\hbar \al$ and $\hbar \be$ stand for the coupling strengths of the Swanson model,  $\al,  \be \in \mathbb{R}$.
To write the Hamiltonian of Eq. (\ref{hswsq}) in coordinate representation, we have to take $\aam$ and $\aap$ as:
\beqn
\aam = \frac{1}{\sqrt{2}} \left ( 
\frac {\hat x} {b_0} + 
{\bf i} \frac {b_0} {\hbar} \hat{p}
\right), ~~~
\aap = \frac{1}{\sqrt{2}} \left ( 
\frac {\hat x} {b_0} - 
{\bf i} \frac {b_0} {\hbar} \hat{p}\right).
\label{hxp0}
\eeqn
In coordinate representation 

\begin{flalign}
&{\mathrm H} (\om,\al,\be)=
\frac 1 2 \hbar (\omega + \alpha +\beta ) \frac {{\hat x}^2} {b_0^2} 
+ \frac 1 2  (\omega - \alpha -\beta ) \frac {b_0^2} {\hbar^2} ~ {\hat p}^2
+  \uni \frac{(\alpha-\beta)}{2} \left \{ \hat x, \hat p\right \}.
\label{hxp}
\end{flalign}
By a suitable transformation, it can be proved that the Hamiltonian of Eq.(\ref{hxp}) is similar to its adjoint \cite{nosswrhs}. In terms of the model space of parameters, different physical regions arise. 
Particularly, if $(\omega^2- 4 \al \be) <0$ and $(\omega-\al-\be) >0$, both $H$ and its adjoint operator are similar to the Hamiltonian of an inverted oscillator. For further details, the reader is kindly referred to \cite{nosswrhs}. 
The mathematical aspects of the study of the inverted oscillator can be found in \cite{chu1,chu2, maru,bermudez}. As reported in \cite{chu1,chu2, maru,bermudez}, the generalized eigenfunctions belong to a Rigged Hilbert space\cite{gelfand,bhom}. 

In what follows, we shall focus on the description of time evolution for the model space parameters  $(\omega^2-4 \al \be)<0$ and $(\omega-\al-\be) >0$. To do so, we shall use the
CSM method  \cite{csm0,csm1,csm2,csm3,csm4,csm5,moiseyev0,moiseyevpr,moiseyev}, which is an alternative approach to deal with the Hamiltonian of Swanson in the broken PT-symmetry phase. 

In the framework of the CSM, we shall introduce the transformation operator
\beqn
{\hat V}(\theta)=
{\rm e}^{\frac{\theta}{2 \hbar}({\hat x}{\hat p}+{\hat p}{\hat x})},
\label{vth}
\eeqn
with a real scaling parameter $\theta$. Under this similarity mapping:

\beqn
\hat V(\theta)^{-1} {\hat x} \hat V(\theta) & = &
\re^{~{\bf i} \theta }~{\hat x},\nonumber \\
\hat V(\theta)^{-1} {\hat p} \hat V(\theta) & = &
\re^{-{\bf i} \theta }~{\hat p}.
\label{eq4}
\eeqn

The Hamiltonian of Eq.(\ref{hxp}) is transformed as ${\mathrm H}(\theta)=\hat V^{-1}(\theta) {\mathrm H} \hat V(\theta)$

\begin{flalign}
 {\mathrm H}(\theta)&={\mathrm H}(\theta,\om,\al,\be) \nonn
 &=
\frac 1 2 \hbar (\omega + \alpha +\beta ) \frac 1 {b_0^2} \re^{2 \uni \th}{\hat x}^2 
+ \frac 1 2  (\omega - \alpha -\beta ) \frac {b_0^2} {\hbar^2} ~ \re^{-2 \uni \th}{\hat p}^2
+  \uni \frac{(\alpha-\beta)}{2} \left \{ \hat x, \hat p\right \}.
\label{htheta}
\end{flalign}
It is straightforward to observe that the adjoint of the operator $H$, $H_c$, is given by 

\beqn
{\mathrm H}_c(\theta)={\mathrm H}(-\theta,\om,\be,\al).
\label{hthc}
\eeqn
Notice that ${\mathrm H}(\theta)$ is invariant under the usual $\pt$-symmetry given by ${\hat x} \rightarrow -\hat x$, ${\hat p} \rightarrow \hat p$, $\uni \rightarrow -\uni$ and $\theta \rightarrow -\theta$.

For $\omega \neq \al+\be$, we can introduce a new set of conjugate operators $\hat X$ and $\hat P$, which obey the usual commutation relation $[\hat X, \hat P]= \uni \hbar$, namely

\beqn
\hat P = \hat  \re^{-\uni \th} p + \uni \hbar \frac{\al -\be}{(\om-\al-\be)b_0^2}  \re^{\uni \th} \hat x, ~~~~~~\hat X= \re^{\uni \th} \hat x.
\label{newxp}
\eeqn 
The Hamiltonian of Eq.(\ref{htheta}), in terms of the operators $\hat P$ and $\hat X$, can be expressed as

\beqn
H= \frac 1{2 m} \hat P^2+ \frac 1 2 k \hat X^2.
\eeqn
We have defined \cite{nosswrhs} $k= m ~ \Omega^2$ and

\beqn
m=m(\omega,\alpha,\beta,b_0)& = \frac{ \hbar}{(\omega-\alpha-\beta) b_0^2}, \nonumber \\
\Omega=\Omega(\omega,\alpha,\beta)& =  \sqrt{\omega^2-4 \alpha \beta}=|\Omega| {\rm {e}}^{{\bf i} \phi}.
\eeqn
As discussed in \cite{nosswrhs}, the sign of $m$ and $\Om^2$ characterise different regions in the parameter model space. We are interested in the study of the region with $m>0$ and $\Om^2<0$. In this case the Hamiltonian $\mathrm H$ is similar to an inverted oscillator.

To obtain the eigenfunctions and the eigenvalues of $\mathrm H$ and $\mathrm H_c$, we shall proceed as follows. 

We shall introduce the operator \cite{noscsm} 

\beqn 
\Upsilon(\theta,x) = {\rm e^{-\frac {\alpha-\beta}{\omega-\alpha-\beta} \frac {{\rm e}^{2 {\bf i} \theta}x^2} {2 b_0^2}}},
\label{opsim}
\eeqn
to perform a similarity transformation on $\mathrm H$:

\beqn
\Upsilon (\theta,x) ~ {\mathrm H}(\theta)   \Upsilon(\theta,x)^{-1}  & = &
{\mathfrak h}(\theta) ,
\label{simih}
\eeqn
where ${\mathfrak h}(\theta)$ is given by
\beqn
{\mathfrak h} (\theta)= \frac{1}{2 m~ }  \left( {\rm e}^{-{\bf i} \theta} \hp \right)^2 +
\frac 1 2 k  \left( {\rm e}^{ {\bf i} \theta} \hx \right)^2.
\label{hosc}
\eeqn

As proved in \cite{noscsm}, for ${\mathrm H}_c(\th)$:

\beqn
\Upsilon(-\th,x)^{-1} {\mathrm H}_c(\th) \Upsilon(-\th,x)& = &   {\mathfrak h}_c (\theta)={\mathfrak h}(-\theta).
\label{simihc}
\eeqn

From Eqs. (\ref{simih}) and (\ref{simihc}), it can be proved that the eigenfunctions of $\mathfrak{h}(\theta)$, $\phi(\theta)$, are related to that of $H$ and $H_c$ as follows: 

\beqn \label{AutHHd}
 \wid{\phi}(\theta,x)&=& \Upsilon(\theta,x)^{-1}~\phi(\theta,x), \nonumber \\
 \ovl{\psi}(\theta,x)&=& \Upsilon(-\theta,x)~\phi(-\theta,x).
\eeqn

As it has been proven in \cite{noscsm}, the Hamiltonians  ${\mathrm H}(\theta)$ and ${\mathrm H}_c(\theta)$ can be related by a similar transformation $S(\theta)$:

\beqn
{\mathrm H}_c(\theta) S(\th)=S(\th) {\mathrm H}(\theta), 
~~~~S(\theta)=\Upsilon(-\theta,x) V(2 \theta) \Upsilon(\theta,x). 
\label{simihhc}
\eeqn 
Thus, $\ovl{\psi}(\theta,x)=S(\th) \wid{\phi}(\theta,x)$.

For further results, the reader is kindly referred to \cite{nosswrhs,noscsm}.



\subsection{Discrete Spectrum.}\label{dis}
The discrete spectrum of  ${\mathrm H}(\theta)$ and ${\mathrm H}_c(\theta)$ and the corresponding eigenfunctions are given by:

\begin{equation}\label{eq 10}
E_n^\pm =\hbar |\Omega| e^{\pm \textbf{i} \frac \pi 2} \left(n+\frac{1}{2}\right)
\end{equation}
\begin{equation}\label{eq 11}
\begin{split}
\tilde{\phi}^\pm_n(\theta,x)&=\Upsilon^{-1}(\theta,x)\phi^\pm_n(\theta,x),~~~\wid{E}_n^\pm=E_n^\pm,\\
\ovl{\psi}^\pm_n(\theta,x)&=\Upsilon(-\theta,x)\left(\phi^\pm_n(\theta,x)\right)^*, ~~~\ovl{E}_n^\pm=E_n^{\pm *},
\end{split}
\end{equation}
where $\phi^\pm_n(\theta,x)$ can be expressed as
\begin{equation}
\phi^\pm_n(\theta,x)=\mathcal{N}_n e^{-e^{2\textbf{i}(\theta\pm \frac \pi 4)}\frac{x^2}{2b_0^2}\vert \sigma\vert^2}H_n\left(e^{\textbf{i}(\theta\pm \frac \pi 4)}\frac{x}{b_0} |\sigma|\right),
\label{phin}
\end{equation}
and
\begin{equation}
    \mathcal{N}_n^{\pm 2}=\dfrac{e^{\textbf{i}(\theta \pm \frac \pi 4)}}{\sqrt{\pi} n! 2^n}\dfrac{|\sigma|}{b_0}.
\end{equation}
We have defined 

\beqn
\sigma= \sqrt{ \frac {m \Om}{\hbar}} b_0.
\label{sgm}
\eeqn

As shown in \cite{noscsm}, the biorthogonal relation reads:

\begin{flalign}
&\int_{-\infty}^{\infty}~({\ovl \psi}^\pm_m(\theta,x))^*\wid \phi^\pm_n(\theta,x) ~{\rm d}x =
\int_{-\infty}^{\infty}~\phi^\pm_m(\theta,x)\phi^\pm_n(\theta,x) ~{\rm d}x
=\delta_{m n}. \nonn
\label{bid}
\end{flalign}

Defining the complex potential $U(\theta,x)=e^{2\textbf{i}(\theta\pm \frac \pi 4)}\frac{x^2}{2b_0^2}\vert \sigma\vert^2$, it can be observed that 
the eigenfunctions $\wid \phi_n(\theta,x)$ are square-integrable, in the sense of Eq.(\ref{bid}), for the  $\theta$-interval where $Re(U(\theta,x))$ is positive.

In the region where $m>0$ and $\Omega^2<0$, we can find two well define $\theta$-domains: 
$I_1=(0,\frac{\pi}{2})$ and $I_2=(\frac{\pi}{2},\pi)$. These intervals are repeated periodically, with period $\pi$.
The eigenfunctions $\wid \phi^+_n(\th,x)$ and $\ovl \psi^+_n(\th,x)$ are square integrable for $\th \in I_2$, while $\wid \phi^-_n(\th,x)$ and $\ovl \psi^-_n(\th,x)$ are square integrable for $\th \in I_1$.

\subsection{Continuum Spectrum}\label{con}

For the sake of completeness, we present the 
eigenfunctions and eigenvalues associated with the continuum spectrum of the Hamiltonians of Eqs. (\ref{htheta}) and (\ref{hthc}). They read

\begin{equation}
\begin{split}
\tilde{\phi}^\pm_E(\theta,x)&=\Upsilon^{-1}(\theta,x)\phi^\pm_E(\theta,x),~~~\wid{E}(\th)=E(\th),\\
\ovl{\psi}^\pm_E(\theta,x)&=\Upsilon(-\theta,x)\left(\phi^\pm_E(-\theta,x)\right), ~~~\ovl{E}(\th)=E(-\th),
\end{split}
\end{equation}
In the previous equation, $E(\th)=\re^{-2 \uni \th}E$, with $E \in \mathbb R$, and $\phi^\pm_E(\theta,x)$ are the eigenfunctions of the continuous spectrum of $\mathfrak{h}(\theta)$:

They are given in terms of the eigenfunctions of $\mathfrak{h}(\theta)$ and $\mathfrak{h}_c(\theta)$, respectively:

\beqn
\mathfrak{h}(\theta) {\phi}^\pm_E(\theta,x) & = & E~\re^{-2 \uni \th}~ {\phi}^\pm_E(\theta,x), \nonn
\mathfrak{h}_c(\theta) {\psi}^\pm_E(\theta,x) & = & E~\re^{2 \uni \th}~ {\psi}^\pm_E(\theta,x),
\eeqn
being ${\psi}^\pm_E(\theta,x)={\phi}^\pm_E(-\theta,x)$, and 

\begin{equation}
\phi^\pm_E(\theta,x)=\mathcal{N}~\Gamma(\nu +1)D_{-\nu-1}\left(\mp\sqrt{-2 \uni}~\re^{\textbf{i} \theta }\vert \sigma\vert \frac{x}{b_0} \right).
\label{phincon}
\end{equation}
In the previous equations, $D_{-\nu-1}(y)$ are the  parabolic cylinder functions, $\nu=-( \uni ~\frac {E~\re^{2 \uni \th}}{\hbar |\Om|} +\frac{1}{2})$, and ${\cal N}$ is the normalisation constant. 

As shown in \cite{noscsm}, the biorthogonality and the completeness relation can be written as

\begin{equation}
    \int_{-\infty}^{\infty} (\bar{\psi}^\pm_E(\theta,x))^*\tilde{\phi}^{\pm}_{E'}(\theta,x) dx =\delta(E-E'),
    \label{bic}
\end{equation}

\begin{equation}
    \sum_{s=\pm}\int_{-\infty}^{\infty} (\bar{\psi}^{-s}_E(\theta,x))^*\tilde{\phi}^{s}_{E}(\theta,x) dE =\delta(x-x').
\end{equation}

In the framework of the CMS, the eigenfunctions are square-integrable, in the sense of Eq. (\ref{bic}), for the particular value of 
$\th= \pi/2$.
\subsection{Time Evolution.}\label{te}

To study the time evolution of different wave packages we shall adopt the formalism of the 
Response Function\cite{response,barton1986}. 

The Response Function of $H(\th)$ and $H_c(\theta)$ can be given in terms of the Response Function of $h(\theta)$ and $h(-\theta)$, respectively. That is, from the equation

\beqn
{\mathfrak h}(\pm \theta) K(\pm \theta,x,x';t) & = & {\bf i} \hbar \frac {\partial  K} {\partial t} (\pm \theta,x,x',t) \nonumber \\
{\displaystyle{\lim}_{t \rightarrow 0^+} }~  K(\th,x,x';t)  & = & \delta(x-x'). 
\label{Kh}
\eeqn
The solution of Eq. (\ref{Kh}) can be found straightforwardly. It reads: 

\begin{flalign}
& K(\th,x,x';t) = \frac {1} {\sqrt{2 \pi {\bf i}\sin (\Omega t)}}  \frac {\sg} { b_0} 
\re^{ {\bf i} \frac {\sg^2 } {2 b_0^2\sin (\Omega t)} \re^{2 \uni \th} \left( (x^2 + x'^2) \cos(\Omega t) - 2 x x'\right)}.
\end{flalign} 

For the system governed by ${\mathrm H}(\theta)$, using \eqref{simih} the Response Function is given by

\begin{flalign}
&\wid K(\th,x,x';t)= \Gamma(\pm \th,x,x') K(\th,x,x',t),
\label{K}
\end{flalign} 
with

\beqn 
\Gamma(\th,x,x')= 
\re^{\frac {\alpha-\beta}{\omega-\alpha-\beta} \re^{2 \uni \th} \frac { x^2-{x'}^2} {2 b_0^2}}.
\label{opsimi}
\eeqn 

While, for ${\mathrm H}_c(\theta)$ the Response Function can be obtained by using \eqref{simihc}, $\ovl K(x,x';t)$ reads:

\begin{flalign}
\ovl K(\th,x,x';t)=\Gamma(-\th,x,x')^{-1} K(-\th,x,x',t).
\label{Kc}
\end{flalign}

Given an initial wave packet $f(x,t_0)$, it evolves in time under ${\mathrm H}(\theta)$ as 

\beqn
\widetilde{f}(\th,x,t)= \int_{-\infty}^{\infty} ~\wid K(x,x';t) \wid f(\th,x',t_0) {\rm{dx'}},
\label{ft}
\eeqn
and under ${\mathrm H}_c(\theta)$ as:
\beqn
\ovl{f}(\th,x,t)= \int_{-\infty}^{\infty} ~\ovl K(\th,x,x';t) \ovl f(\th,x') {\rm{dx'}}.
\label{ftc}
\eeqn
with 

\beqn
\wid f(\th,x,0) & = & \Upsilon(\th, x)^{-1} \re^{\uni \frac \th 2} f(\re^{\uni \th}  x), \nonn
\ovl f(\th,x,0) & = & \Upsilon(-\th, x) \re^{- \uni \frac \th 2} f(\re^{-\uni \th}x). \nonn
\label{inicondition}
\eeqn
The normalization condition is given by

\beqn
\int_{-\infty}^{\infty} ~ \ovl f(\th,x,t)^* \wid f(\th,x,t){\rm {dx}}=1 \label{norma1}.
\eeqn

In the region of consideration, $m>0$ and $\Om^2<0$:

\begin{flalign}
&K(\th,x,x';t)= -\frac {1} {\sqrt{2 \pi {\bf i}\sinh (|\Omega| t)}}  \frac {|\sg|} { b_0} \re^{-\uni \frac \pi 4}
\re^{ {\bf i}  \frac {|\sg|^2 } {2 b_0^2\sinh (|\Omega| t)} \re^{2\uni \th} \left( (x^2 + x'^2) \cosh(|\Omega| t) - 2 x x'\right)}.
\end{flalign} 

The time evolution of the discrete eigenstates of ${\mathrm H}$ and ${\mathrm H}_c$ can be obtained from Eqs. (\ref{ft}) and (\ref{ftc}). The details are given in \ref{ap1}.
After some algebra, the time evolution of the eigenstates of $\mathrm H$ is given by:

\begin{flalign}
& \wid \phi_n^\pm(\th,x,t)= \mp \re^{ - \uni   \frac \pi 2 n} \re^{ \pm n  |\Om| t} \wid \phi_n^\pm(\th,x) 
= \mp \re^{ - \uni \frac \pi 2 n} \re^{ \pm n  |\Om| t} \Upsilon(\th,x)^{-1} \re^{\uni \frac \th 2} \phi_n^\pm(\th,x),\nonn
\end{flalign}
and those of ${\mathrm H}_c$ read:
\begin{flalign}
& \ovl \psi_n^\pm(\th,x,t)= \pm \re^{  \uni   \frac \pi 2 n} \re^{ \mp n  |\Om| t} \ovl \psi_n^\pm(\th,x)
= \pm \re^{  \uni   \frac \pi 2 n} \re^{ \mp n  |\Om| t} \Upsilon(-\th,x)  ( \re^{\uni \frac \th 2} \phi_n^\pm(\th,x))^*,\nonn
\end{flalign}

In the framework of the CSM, the time evolution of a given initial state, $f(x)$, is split into a retarded state ($t>t_0$) and an advanced state ($t<t_0$) \cite{hatano}. 
For the time evolution under the action of $H(\th)$, it reads
\beqn
\wid f  (\th,x,t)  & = &  \wid f^+ (\th,x,t) + \wid f^- (\th,x,t), \nonn 
\wid f^\pm(\th,x,t)  & = & \sum_n ~\wid c_n^\pm (t_0) ~\wid \phi_n^\pm(\th,x,t) ,  
\eeqn
with $\wid c_n^\pm (t_0) =\langle \ovl \psi_n^\pm (\th,x)|f (\th,x,t_0)\rangle$. Similarly, for $H_c(\th)$, we have

\beqn
\ovl f  (\th,x,t)  & = & \ovl f^+ (\th,x,t) + \ovl f^- (\th,x,t), \nonn
    \ovl f^\pm(\th,x,t)  & = & \sum_n ~ \ovl c_n^\pm (t_0) ~\ovl \psi_n^\pm(\th,x,t), 
\eeqn
with $\ovl c_n^\pm (t_0) =\langle \wid \phi_n^\pm (\th,x)|f (\th,x,t_0)\rangle$.

\subsection{Wigner Function.}\label{wig}

We shall define the Complex Scaling Wigner Function of $f(x,t)$ as:
\beqn  
W(\th,x,p,t) & = & \int_{-\infty}^{\infty}~\ovl f\left( \th,x+ \frac y 2,t \right)^*~\wid{f} \left(\th,x-\frac y 2,t\right)~\re^{\uni \frac{p} {\hbar }y }~{\rm {dy}} \nonn
& = & \sum_{m,n}~\ovl c_n^\pm (t_0)^* \wid c_n^\pm (t_0) W_{m n}^\pm(\th,x,p),
\label{wigf}
\eeqn
where $W_{m n}^\pm(\th,x,p,t)$ is the Wigner function associated with the eigenfunctions of $H(\th)$:

\begin{flalign} 
W_{m n}^\pm(\th,x,p,t)=\int_{-\infty}^{\infty}~\ovl{\psi}_m^\pm \left( \th,x+\frac y 2,t \right)^*~\wid{\phi}_n^\pm \left(\th,x-\frac y 2,t\right)~\re^{\uni \frac{p} {\hbar }y }~{\rm {dy}}.
\end{flalign}
The explicit calculation of $W_{m n}^\pm(\th,x,p)$ can be found in \ref{ap1}. After some algebra, it reads 
\beqn
W_{m n}^-(u,v) & = & 
2 \sqrt{\frac{m!}{n!}} \re^{- (n-m)|\Om|t}\re^{-u v} u^{n-m}{\rm L }_m^{n-m}(~2 u v), ~~~ m<n,\nonn
& = &
2 \sqrt{\frac{n!}{m!}} \re^{- (n-m)|\Om|t}\re^{-u v} v^{m-n}{\rm L }_n^{m-n}(~2 u v), ~~~ n<m ,\nonn
& = &
2 \re^{-u v} {\rm L }_n(~2 u v), ~~~ n=m ,\nonn
W_{m n}^+(u,v) & = & 2 \sqrt{\frac{m!}{n!}} \re^{~(n-m)|\Om|t}\re^{~ u v} v^{n-m}{\rm L }_m^{n-m}(-2 u v), ~~~ m<n,\nonn
& = & 2 \sqrt{\frac{n!}{m!}} \re^{~(n-m)|\Om|t}\re^{~ u v} u^{m-n}{\rm L }_n^{m-n}(-2 u v),~~~ n<m ,\nonn
& = &
2 \re^{~u v} {\rm L }_n(-2 u v), ~~~ n=m ,\nonn
u (\th,x,p) & = &  \lpth-\frac {\alpha-\beta}{(\omega-\alpha-\beta)|\sg|^2} \lth -\uni \lth, \nonn
v (\th,x,p) & = &  \lpth-\frac {\alpha-\beta}{(\omega-\alpha-\beta)|\sg|^2} \lth +\uni \lth. \nonn
\label{wigmn}
\eeqn

From the Wigner Function, the package density is given by

\beqn
\rho(\th,x,t)= \int_{-\infty}^{\infty}~ W( \th,x,p,t)~{\rm {dp}}.
\label{pack}
\eeqn

Given the $\rho(\th,x,t)$ the equation of continuity equation, see \ref{ap2}, can be written as 

\beqn
\partial_t (\rho(\th,x,t))+\partial_x J(\theta,x,t)=0,
\eeqn
with the current density computed as 

\begin{flalign}
& J(\theta,x,t)= \frac {\hbar}{2 m \uni} \re^{-2 \uni \th } 
\left( 
\ovl{f}^*(\theta,x,t)  \left( \frac{{\partial } } {\partial x}{\wid{f}} (\theta,x,t)\right )-
\left(\frac{{\partial } } {\partial x}  {\ovl{f}}^* (\theta,x,t)\right) \wid{f}(\theta,x,t) 
\right). \nonn
\end{flalign}

In terms of the package density, the probability of persistence in the barrier can be computed as
 
\begin{flalign}
& Q(\th,L, t)= \int_{-L}^{L}~ \rho(\th,x,t) {\rm {dx}}.
\end{flalign}


To complete the analysis of the problem, we shall compute the survival probability of an initial package as

\begin{flalign}
& {\cal P}_s(\th,t) = \bigg| \int_{-\infty}^{\infty}~  \ovl f (\th,x,t_0)^*  \wid f(\th,x, t) {\rm {dx}} ~\bigg|^2.\label{sur}
\end{flalign} 

\section{Results and Discussion}\label{results}

The implementation of the CSM to describe experimental results \cite{csm4,csm5} requires fixing the value of the parameter of $\theta$ phenomenologically from the available data. In what follows, we shall give our results in function of $\th$ and we shall fix $\th$ to the value $\pi/4$ to reproduce the results of the Swanson model of Eq.(\ref{hswsq}).

Before discussing the time evolution of a particular wave package, let us analyse the behaviour of the Wigner functions, $W_{m,n}^{-} (\th,x,p)$, associated with the eigenstates of the problem. We performed the study in terms of the asymmetry of the coupling constants $\al$ and $\be$ for different values of $|\Om|$. To simplify the study, we shall take the diagonal components of $W_{m,n}^{-}(\th,x,p)$. 

\begin{figure}
\includegraphics[width=12cm]{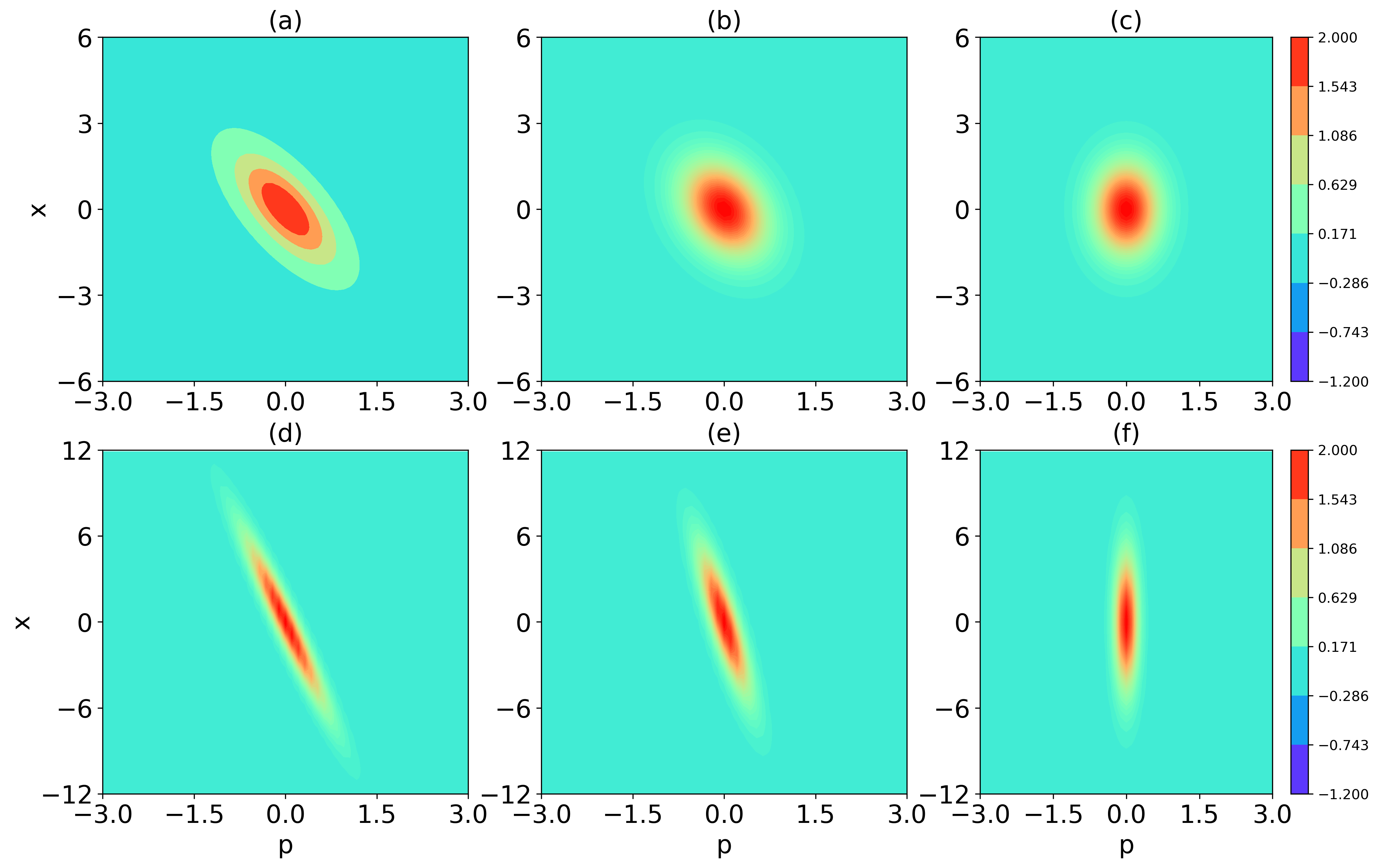}
\caption {Behaviour of the diagonal elements of $W_{m n}^{-}(\th,x,p)$ of Eq. (\ref{wigmn}) for different values of the coupling constants $\alpha$ and $\beta$, at $\om=1$ and $\th=\pi/4$.  The Figure displays the results obtained for $n=m=0$. Panels (a), (b) and (c) correspond to values of $|\Om|=1$, with $\al=-2,~\be=-1/4$, $\al=-1,~\be=-1/2$ and $\al=-\sqrt{1/2},~\be=-\sqrt{1/2}$, respectively.  In Panels (d),(e) and (f) we display the results obtained for $|\Om|=0.1$, with $\al=-2,~\be=-101/800$, $\al=-1,~\be=-101/400$ and $\al=-\sqrt{101}/20,~\be=-\sqrt{101}/20$, respectively.}
\label{fig1}
\end{figure}

In Figure \ref{fig1}, we have depicted the results obtained for $W_{0 0}^{-}(\th,x,p)$ of Eq. (\ref{wigmn}) for different values of the coupling constants $\alpha$ and $\beta$, at $\om=1$ and $\th=\pi/4$.  The Figure displays the results obtained for $n=m=0$. Panels (a), (b) and (c) correspond to values of $|\Om|=1$, with $\al=-2,~\be=-1/4$, $\al=-1/,~\be=-1/2$ and $\al=-\sqrt{1/2},~\be=-\sqrt{1/2}$, respectively.  In Panels (d),(e) and (f) we display the results obtained for $|\Om|=0.1$, with $\al=-2,~\be=-101/800$, $\al=-1,~\be=-101/400$ and $\al=-\sqrt{101}/20,~\be=-\sqrt{101}/20$, respectively. For the state with $n=m=0$, the Wigner Functions take positive values. 

\begin{figure}
\includegraphics[width=12cm]{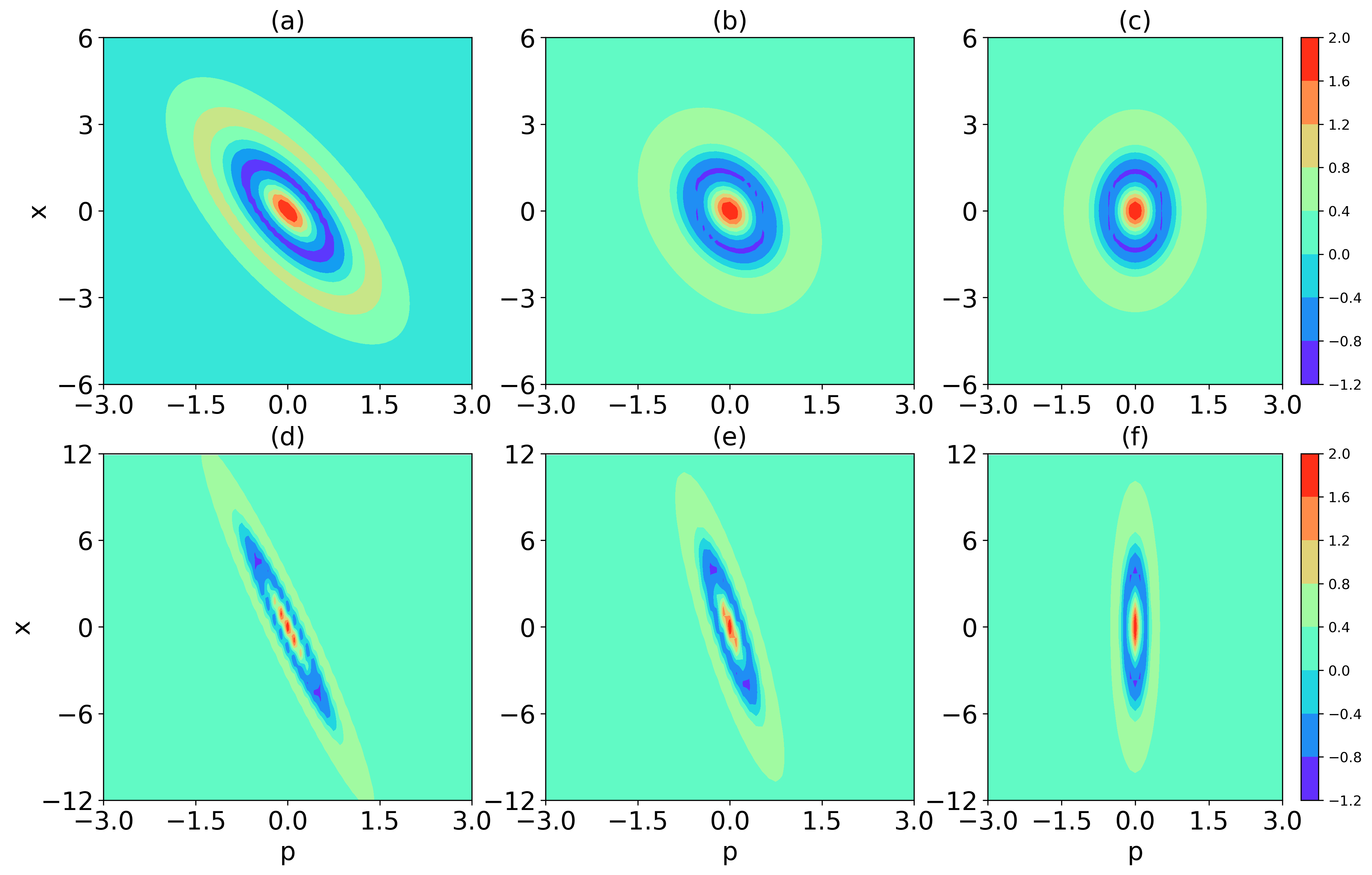}
\caption {Behaviour of the diagonal elements of $W_{m n}^{-}(\th,x,p)$ of Eq. (\ref{wigmn}) for different values of the coupling constants $\alpha$ and $\beta$, at $\th=\pi/4$.  The Figure displays the results obtained for $n=m=2$. Panels (a), (b) and (c) correspond to values of $|\Om|=1$, while Panels (d),(e) and (f) correspond to values of $|\Om|=0.1$. The values for the coupling constants are the same as those of Figure \ref{fig1}.}
\label{fig2}
\end{figure}

Figure \ref{fig2} shows the results for the state corresponding to $n=2$, the coupling constants are the same as those of Figure\ref{fig1}. For these states, and generally for states with $n \neq 0$, the Wigner Functions can take negative values. 

It can be observed that the Wigner function is stretched, along the line

\beqn 
x& \approx & \left( \frac{\gamma^2 +|\sg|^4-1}{2 \gamma} +\sqrt{1+ \left(\frac{\gamma^2 +|\sg|^4-1}{2 \gamma} \right)^2} \right) \frac{p}{\hbar}, \nonn
\gamma & = & \frac{\al-\be}{\om-\al-\be},
\eeqn
and that, in both cases, for $\Om$ close to zero, the Wigner functions are stretched along the $x$-axe.

The behaviour at $\Om$ close to value zero, can be understood as a consequence of being in the proximity of an Exceptional Point (EP). 
For the present model, at fixed $\om$, EPs appear at values of $\al$ and $\be$ such that $\Om=0$ \cite{noscsm}. The consequence of being in the vicinity of EPs will be discussed in the following examples.

Let us discuss the time evolution, for $t>0$, of the following wave packages \cite{hatano}: 

\begin{flalign}
&f_c(x,0)={\cal N}_c ~
\re^{-\re^{2\uni \left(\th-\frac \pi 4 \right)} \frac {|\sg|^2 }{2 b_0^2} ~x^2 }\cosh \left( 2~\re^{\uni (\th-\pi/4)} \frac {|\sg|}{b_0}  x\right),
~~~{\cal N}_c= \left( \frac{\re^{\uni \left(\th- \frac \pi 4 \right)}}{\pi^{1/2}\re^2 \cosh(2)} \frac {|\sg|}{b_0} \right)^{1/2}
,\label{fc} \\
&f_s(x,0)={\cal N}_s ~
\re^{-\re^{2\uni \left(\th-\frac \pi 4 \right)} \frac {|\sg|^2 }{2 b_0^2} ~x^2 }\sinh \left( 2~\re^{\uni (\th-\pi/4)} \frac {|\sg|}{b_0}  x\right),
~~~{\cal N}_s= \left( \frac{\re^{\uni \left(\th- \frac \pi 4 \right)}}{\pi^{1/2}\re^2 \sinh(2)} \frac {|\sg|}{b_0} \right)^{1/2}
,\label{fs} \\
&f_g(x,0)={\cal N}_g ~\re^{-\left (\re^{\uni \left(\th- \frac \pi 4 \right)} \frac {|\sg|}{b_0}  x-2 \right)^2/2},
~~~{\cal N}_g= \left( \frac{\re^{\uni \left(\th- \frac \pi 4 \right)}}{\pi^{1/2}} \frac {|\sg|}{b_0} \right)^{1/2}.
\label{fg} 
\end{flalign}
The time evolution of these wave packages is given by:

\beqn
\wid f_c^{-}(y,t)&= \Upsilon^{-1}(y)~\sum_{n=0}^{\infty} ~ \chi_c(2n)~\re^{-\left(2 n +\frac 12 \right)|\Om| t}~\re^{-\frac{y^2}{2} } H_{2 n}(y),~~
& \chi_c(n)= \frac {1} {n!} \re {\cal N}_c \nonn 
& =  \Upsilon^{-1}(X)~\frac {\re^{-\frac{y^2}{2} } \cosh(2 \re^{-|\Om| t} y)}  {\sqrt{\pi^{1/2} \cosh(2)} } \re^{-\frac 12 |\Om| t} \re^{-(\re^{-2 |\Om| t} )},& \nonn
\wid f_s^{-}(y,t) & = \Upsilon^{-1}(y)~\sum_{n=0}^{\infty} ~ \chi_s(2 n+1) ~\re^{-\left(2 n+\frac 32 \right) |\Om| t}~\re^{-\frac{y^2}{2} } H_{2 n+1}(y),~~
& \chi_s(n)=\frac {1 } {n!} \re {\cal N}_s ,\nonn
& = \Upsilon^{-1}(X)~\frac {\re^{-\frac{y^2}{2} } \sinh(2 \re^{-|\Om| t} y)}  {\sqrt{\pi^{1/2} \sinh(2)} } \re^{-\frac 12 |\Om| t}
\re^{-(\re^{-2 |\Om| t} )}, & \nonn
\wid f_g^{-}(y,t) & = \Upsilon^{-1}(X)~\sum_{n=0}^{\infty} ~\chi_g(n)~\re^{-\left(n +\frac 1 2\right)|\Om| t}~\re^{-\frac{y^2}{2} } H_{n}(y), ~~
&\chi_g(n)=\frac {1 } {n!} \re^{-1} {\cal N }_g,\nonn
& =  \Upsilon^{-1}(X)~\frac {\re^{-\frac{(y-2 \re^{-|\Om| t})^2}{2} } }  {\pi^{1/4} \re^{(1-\re^{-2 |\Om| t} )}} \re^{-\frac 12 |\Om| t}, & 
\eeqn
with $y=\re^{\uni (\th-\frac \pi 4) }x \frac {|\sigma|} {b_0}=\re^{-\uni \frac \pi 4}X \frac {|\sigma|} {b_0}$. The expansions are valid for $0< \th \le \pi/2$.

Taken into account Eq.(\ref{inicondition}), it can be shown that $\ovl f_i^{-}(y,t)$ ($i=c,s,g$), is given by

\beqn
\ovl f_c^{-}(y,t) 
 & = & \Upsilon(X^*)~\sum_{n=0}^{\infty} ~ \chi_c(2 n)~\re^{\left(2 n +\frac 12 \right)|\Om| t}\re^{-\frac{y^{*2}}{2} } H_{2 n}(y^*) \nonn 
 & = & \Upsilon(X^*)~\frac {\re^{-\frac{y^{*2}}{2} } \cosh(2 \re^{|\Om| t} y^*)}  {\sqrt{\pi^{1/2} \cosh(2)}} \re^{\frac 12 |\Om| t}
 \re^{(\re^{2 |\Om| t} )}
 , \nonn
 \ovl f_s^{-}(y,t) 
& = & \Upsilon(X^*)~\sum_{n=0}^{\infty} ~ \chi_2(2 n+1)~\re^{\left(2 n +\frac 32 \right)|\Om| t}~\re^{-\frac{y^{*2}}{2} } H_{2 n+1}(y^*),\nonn 
 & = & \Upsilon(X^*)~\frac {\re^{-\frac{y^{*2}}{2} } \sinh(2 \re^{|\Om| t} y^*)}  {\sqrt{\pi^{1/2} \sinh(2)} } \re^{\frac 12 |\Om| t}
 \re^{(\re^{2 |\Om| t} )} 
 ,  \nonn
\ovl f_g^{-}(y,t) 
& =& \Upsilon(X^*)~\sum_{n=0}^{\infty} ~\chi_g(n)~\re^{\left( n +\frac 12 \right)|\Om| t}\re^{-\frac{y^{*2}}{2} } H_{n}(y^*),\nonn
 & = & \Upsilon(X^*)~\frac {\re^{-\frac{(y^*-2 \re^{|\Om| t})^2}{2} } }  {\pi^{1/4} \re^{(1-\re^{2 |\Om| t} )}}\re^{\frac 12 |\Om| t},
\eeqn

\begin{figure}
\includegraphics[height=0.9\textheight]{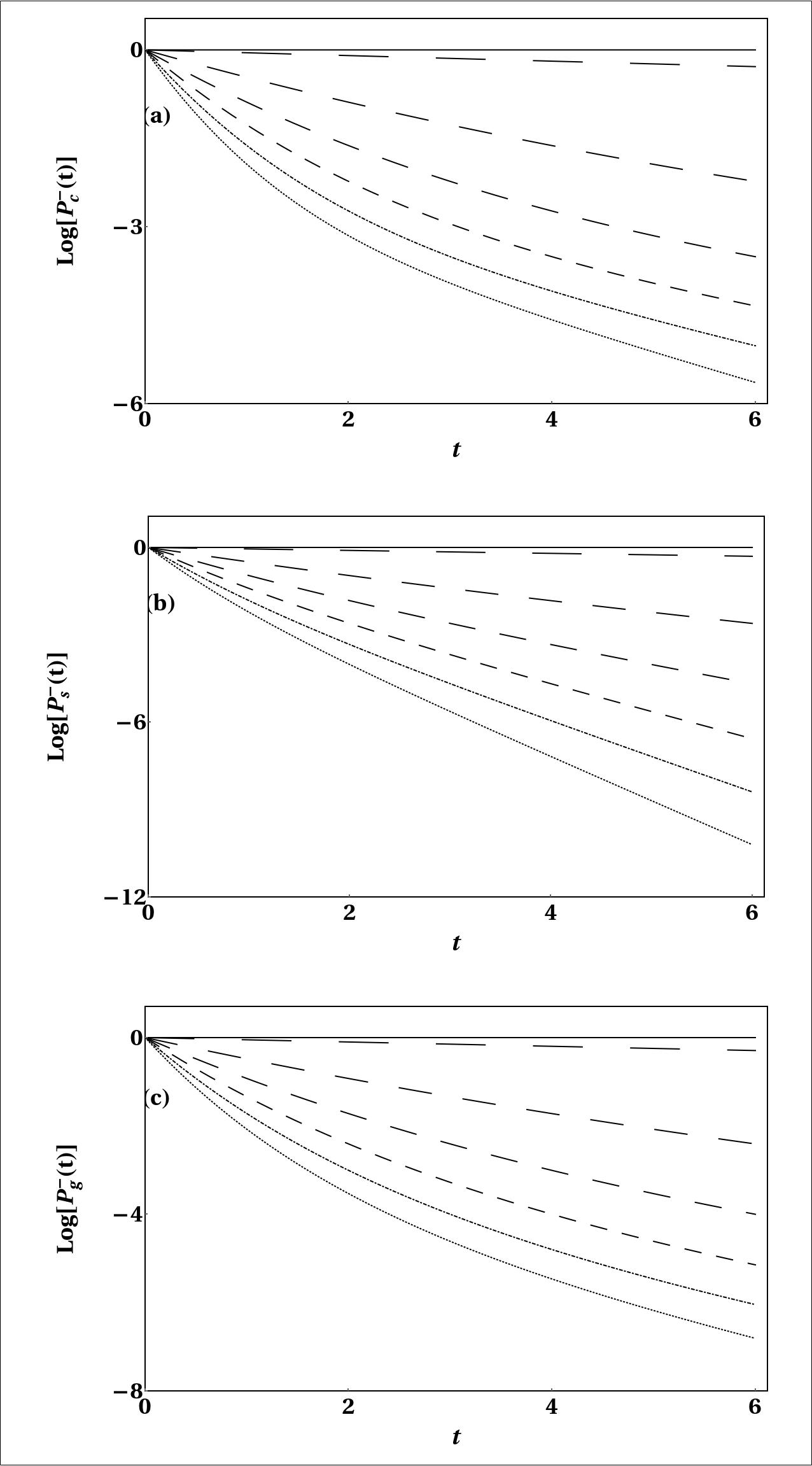}
\caption {Logarithm of the Survival Probabilities as a function of time of the different wave packages. In Panels (a), (b) and (c), we present the results obtained for the wave packages of Eqs. (\ref{fc}) (\ref{fs}) and (\ref{fg}), respectively. The time is given in $\hbar |\Om|$ units. From bottom to top, $|\Om|=0.5,~0.4,~0.3,~0.2,~0.1,~0.01$.}
\label{fig3}
\end{figure}

From Eq. (\ref{sur})one can calculate the survival probability associated with each wave package. They read

\begin{flalign}
& {\cal P}_c^{-}(t) =  \bigg | \frac {\cosh(2 \re^{- |\Om| t})} {\cosh(2)} \re^{-\frac 12 |\Om| t} \bigg |^2,\nonn
& {\cal P}_s^{-}(t) =  \bigg | \frac {\sinh(2 \re^{- |\Om| t})} {\sinh(2)} \re^{-\frac 12 |\Om| t} \bigg |^2,\nonn
& {\cal P}_g^{-}(t) =  \bigg | \re^ { 2( \re^{- |\Om| t}-1)} \re^{-\frac 12 |\Om| t}  \bigg |^2.
\end{flalign}

Figure \ref{fig3} depicts the results obtained for the Logarithm of the Survival Probabilities of the different packages as a function of the time. In Panels (a), (b) and (c), we present the results obtained for the wave packages of Eqs. (\ref{fc}), (\ref{fs}) and (\ref{fg}), respectively. The time is given in $\hbar |\Om|$ units. From bottom to top, $|\Om|=0.5,~0.4,~0.3,~0.2,~0.1,~0.01$. Notice that, as $|\Om|$ is close to its value at the EPs, $|\Om|=0$, the Survival Probability approaches the value $1$. That is the wave packages remain unchanged at the EPs. 

We have computed the Wigner Functions of Eq.(\ref{wigf}) associated with the different packages. They read

\begin{flalign}
& W^-_g ({\cal P},{\cal X})= 
\frac 1 \pi 
\re^{-({\cal P}-2~\uni \sinh(|\Om| t))^2- ({{\cal X}} -2 \cosh(|\Om| t))^2}, \nonn
& W^-_c ({\cal P},{\cal X})= 
\frac {\re^{-{\cal P}^2- {\cal X}^2 } } {2 \pi \cosh (2 )} 
\left( 
\re^2 
\cosh ( 4 {\cal X} \sinh( |\Om| t)+4 \uni {\cal P} \cosh( |\Om| t)) + \right. \nonn
& ~~~~~~~~~~~~~~~~~~~~~~~~~~~~~~~~~~~~~~~~~ \left. 
\re^{-2} 
\cosh ( 4 {\cal X} \cosh( |\Om| t)+4 \uni {\cal P} \sinh( |\Om| t))
\right) 
\nonn
& W^-_s ({\cal P},{\cal X})= 
\frac {\re^{-{\cal P}^2- {\cal X}^2 } } {2 \pi \sinh (2 )} 
\left( 
-\re^2 
\cosh ( 4 {\cal X} \sinh( |\Om| t)+4 \uni {\cal P} \cosh( |\Om| t)) + \right. \nonn
& ~~~~~~~~~~~~~~~~~~~~~~~~~~~~~~~~~~~~~~~~~ \left. 
\re^{-2} 
\cosh ( 4 {\cal X} \cosh( |\Om| t)+4 \uni {\cal P} \sinh( |\Om| t))
\right) 
\nonn
\end{flalign}
where to simplify the notation, we have defined

\begin{flalign}
& {\cal P} = \lpth - \frac{\al -\be}{(\om-\al-\be)|\sg|^2} \lth ,\nonn
& {\cal X} = \lth.
\end{flalign}

\begin{figure}
\includegraphics[width=12cm]{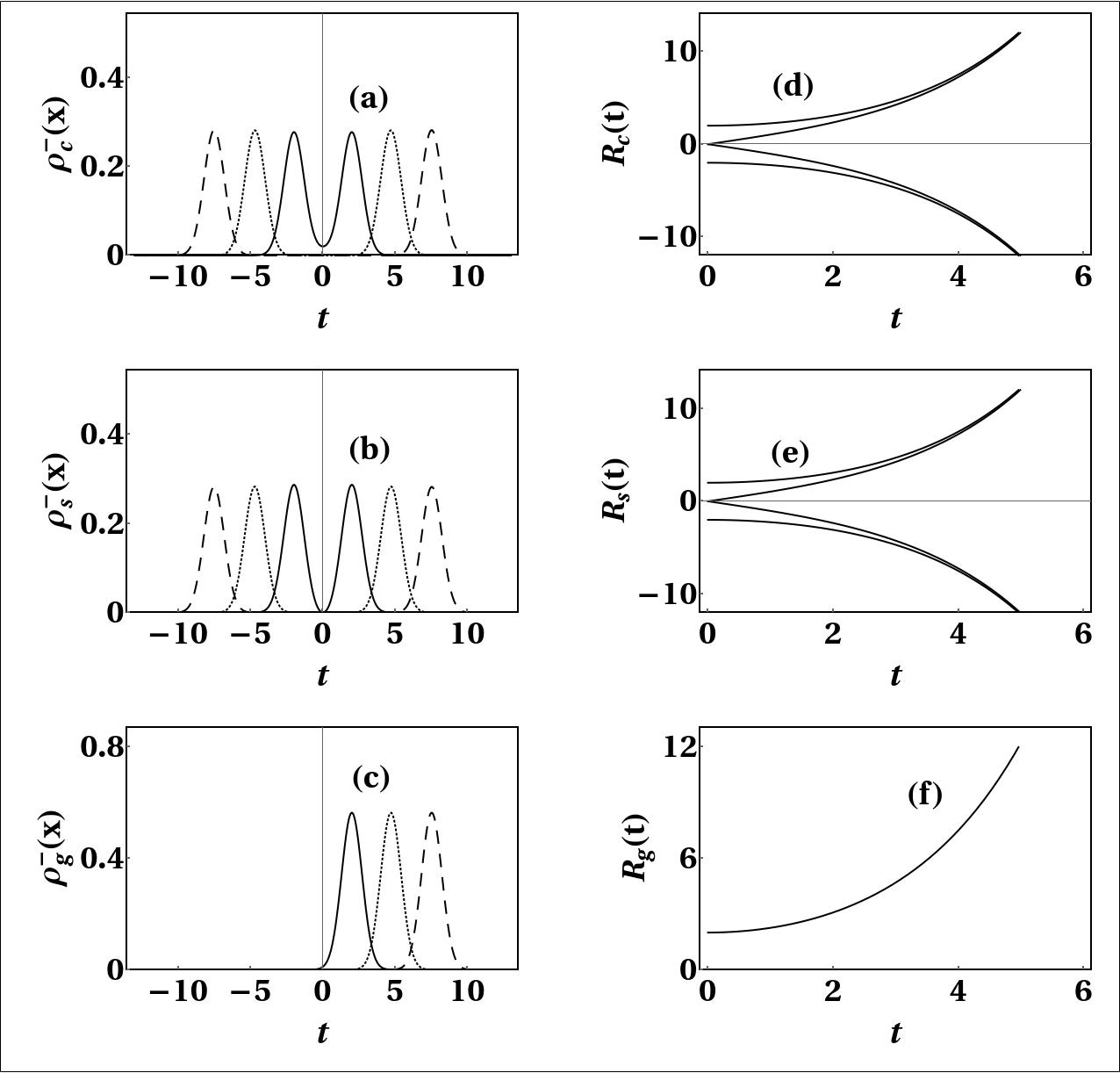}
\caption {Density function for the wave package as a function of time of the position, for different instants of time, for the case $\th=\pi/4$. In Panels (a), (b) and (c), we present the results obtained for the cases presented in Eqs. (\ref{rhoc1}), (\ref{rhos1}), and (\ref{rhog1}), respectively. Solid, Dashed and Long-Dashed Lines correspond to $t=0,~3,~4$, $t$ is given in $\hbar |\Om|$ units. Panels (d), (e), and (f) show the behaviour of the centres of the wave packages of Panels (a), (b) and (c), respectively.}
\label{fig4}
\end{figure}

From the Wigner Functions, it is straightforward to compute the different density distributions, both in $\cal{X} $ and ${\cal P} $. We can write them as:

\begin{eqnarray}
\rho_c^-({\cal X},t)
& = 
& \left(
\re^{2} 
\left( 
\re^{-({\cal X}- 2 \cosh(|\Om| t))^2}+ \re^{-({\cal X}+ 2 \cosh(|\Om| t))^2}
\right)
+\right. \nonn
 & & ~~~\left . \re^{-2} 
\left(
\re^{-({\cal X}- 2 \sinh(|\Om| t))^2} + \re^{-({\cal X}+ 2 \sinh(|\Om| t))^2}
\right)
\right) \frac{1}{4 \sqrt{\pi} \cosh(2)}, \nonn
& &\label{rhoc1}
\\
\rho_s^-({\cal X},t)
& = 
& \left(
\re^{2} 
\left( 
\re^{-({\cal X}- 2 \cosh(|\Om| t))^2}+ \re^{-({\cal X}+ 2 \cosh(|\Om| t))^2}
\right)
- \right. \nonn
 & & ~~~\left. \re^{-2} 
\left(
\re^{-({\cal X}- 2 \sinh(|\Om| t))^2} + \re^{-({\cal X}+ 2 \sinh(|\Om| t))^2}
\right)
\right) \frac{1}{4 \sqrt{\pi} \sinh(2)},
\nonn
& &
\label{rhos1}
\\
\rho_g^-({\cal X},t) & = & \re^{-({\cal X}-2 \cosh(|\Om| t))^2} \frac{1}{{\sqrt{\pi }}},
\label{rhog1}
\end{eqnarray}

After a large interval of time ($|\Om| t >> 0$), it is simple to prove that

\begin{eqnarray}
\rho_c^-({\cal X},t)
& \rightarrow  
& \left(  
\re^{-({\cal X}- \re^{|\Om| t})^2}+ \re^{-({\cal X}+ \re^{|\Om| t})^2}
\right)
\frac{1}{2 \sqrt{\pi}},
\label{rhoc2}
\\
\rho_s^-({\cal X},t)
& \rightarrow  
& \left(  
\re^{-({\cal X}- \re^{|\Om| t})^2}+ \re^{-({\cal X}+ \re^{|\Om| t})^2}
\right)
\frac{1}{2 \sqrt{\pi}},
\label{rhos2}
\\
\rho_g^-({\cal X},t) & \rightarrow & \re^{-({\cal X}- \re^{|\Om| t})^2} \frac{1}{{\sqrt{\pi }}},
\label{rhog2}
\end{eqnarray}

The mean value of the operators associated with ${\cal X}$ are given by

\begin{equation}
\begin{array}{|l|l|l|l|}
\hline
& ~~~~~ \langle {\cal X}\rangle & ~~~~~~~~~~~~~~ \langle {\cal X}^2 \rangle & ~~~~~~~~~~~~~~ \Delta ^2 {\cal X} \\
\hline 
g &   2 \cosh ( |\Om| t) & 1/2 + 4 \cosh (|\Om| t)^2 &1/2  \\ 
c &0  & 1/2 +2 \tanh( 2)+ 2 \cosh (2 |\Om| t) & 1/2 +2 \tanh( 2)+ 2 \cosh (2 |\Om| t)  \\ 
s &0  & 1/2 +2 \coth( 2)+ 2 \cosh (2 |\Om| t) & 1/2 +2 \coth( 2)+ 2 \cosh (2 |\Om| t)  \\ 
\hline
\end{array}
\end{equation}

A particular case is obtained by taking $ \th=\pi/4$  so that the variable ${\cal X} \in \mathbb{R}$. 

In Figure \ref{fig4}, we display the evolution of density spatial distribution for the different wave packages. We have taken $\th=\pi/4$. In Panels (a), (b) and (c), we present the results obtained for the cases presented in Eqs. (\ref{rhoc1}), (\ref{rhos1}), and (\ref{rhog1}), respectively. Solid, Dashed and Long-Dashed Lines correspond to $t=0,~3,~4$, $t$ is given in $\hbar |\Om|$ units. Panels (d), (e), and (f) show the behaviour of the centres of the wave packages of Panels (a), (b) and (c), respectively. From Eqs. (\ref{rhoc1}) and  (\ref{rhos1}), it can be noticed that the waves packages evolve as the superposition of four different packages, $R_c(t)$ and $R_s(t)$ take the values $\pm 2 \sinh(|\Om| t)$ 
and $\pm 2 \cosh(|\Om| t)$. Meanwhile, for (\ref{rhog1}), $R_g(t)$ takes the value $2 \cosh(|\Om| t)$. The different packages evolve as travelling waves. 

The Exceptional Points of the problem are obtained for the set of model parameters obeying $\om^2=4 \al \be$ \cite{nosswrhs,noscsm}, this means $\Om=0$. The survival probability is equal to 1 for all the packages we have analysed, and the spatial density distributions, at all instants of time, are given by

\begin{eqnarray}
\rho_c^-({\cal X})
& = & 
\frac{\cosh^2(2 {\cal X}) \re^{-{\cal X}^2}}{\re^2 \sqrt{\pi} \cosh(2)}, \nonn
\rho_c^-({\cal X})
& = & 
\frac{\sinh^2(2 {\cal X}) \re^{-{\cal X}^2}}{\re^2 \sqrt{\pi} \sinh(2)}, \nonn
\rho_g^-({\cal X}) & = & \re^{-({\cal X}- 2)^2} \frac{1}{{\sqrt{\pi }}}.
\label{rhog3}
\end{eqnarray}

\begin{figure}
\includegraphics[width=14cm]{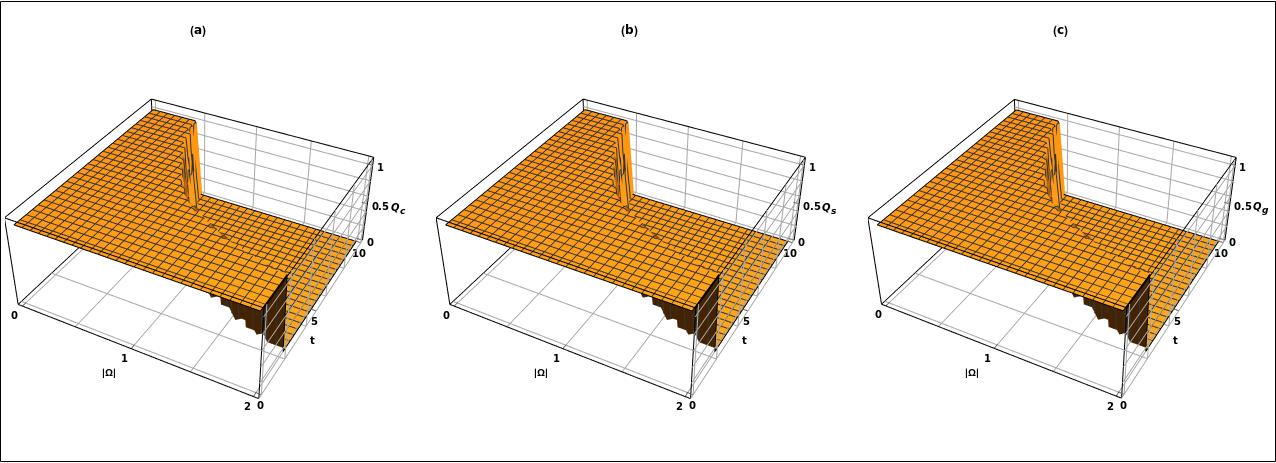}
\caption {Probability of Persistence, for the different wave packages as a function of the time and the frequency,$\{t,|\Om|\}$, for the case $\th=\pi/4$. In Panels (a), (b) and (c), we present the results obtained with $L=200$, for $Q_c(L,|\Om|,t)$, $Q_s(L,|\Om|,t)$ and $Q_g(L,|\Om|,t)$, respectively. $t$ is given in units of $\hbar |\Om|$.}
\label{fig5}
\end{figure}



The probability of persistence of the different packages is given by:

\begin{flalign}
& Q_g(L,\Om,t) = \frac 1 2 \left(
{\rm{Erf}} \left(\llth+2 \cosh(\Omt)\right)+{\rm{Erf}} \left(\llth-2 \cosh(\Omt)\right)\right), \nonn
& Q_c(L,\Om,t) = \frac {\re^{2}} {4 \cosh(2)} \left(
{\rm{Erf}} \left(\llth+2 \cosh(\Omt)\right)+{\rm{Erf}} \left(\llth-2 \cosh(\Omt))\right)
\right) +\nonn
&~~~~~~~~~~~~~~\frac {\re^{-2}} {4 \cosh(2)} \left(
{\rm{Erf}} \left(\llth+2 \sinh(\Omt)\right)+{\rm{Erf}} \left(\llth-2 \sinh(\Omt))\right)
\right) \nonn
& Q_s(L,\Om, t) = \frac {\re^{2}} {4 \sinh(2)} \left(
{\rm{Erf}} \left(\llth+2 \cosh(\Omt)\right)+{\rm{Erf}} \left(\llth-2 \cosh(\Omt))\right)
\right) -\nonn
&~~~~~~~~~~~~~~\frac {\re^{-2}} {4 \sinh(2)} \left(
{\rm{Erf}} \left(\llth+2 \sinh(\Omt)\right)+{\rm{Erf}} \left(\llth-2 \sinh(\Omt))
\right)
\right) \nonn
\label{fgaussok}
\end{flalign}
where $Erf(z)$ stands for the error function of argument $z$. As the wave packages evolve with time, the values of $L$ (in units of $b_0/|\sg|$) must be of the order of $\re^{|\Om| t}$ to guarantee permanence in the barrier.

In Figure \ref{fig5}, we plot the Probability of Persistence, for the different wave packages as a function of the time, $t$, and the frequency,$|\Om|$, for the case $\th=\pi/4$. In Panels (a), (b) and (c), we present the results obtained with $L=200$, for $Q_c(L,|\Om|,t)$, $Q_s(L,|\Om|,t)$ and $Q_g(L,|\Om|,t)$, respectively. $t$ is given in units of $\hbar |\Om|$. At fixed values of $L$, the persistence in the barrier, as a function of time, increases when the values of $|\Om|$ decrease and as they approach the vicinity of the EPs. Also, as the behaviour of the Probability of Persistence in the barrier is dominated by $\re^{|\Om| t}$, there are no considerable differences in the behaviour among the packages, as observed from the Figure.

The density current for the previous packages is given by

\begin{flalign}
& J_c({\cal X},t) = \frac{\hbar}{m} \frac{|\sigma|}{b_0}
\frac {\re^{- \uni (\th-\pi/4)} \re^{-({\cal X}^2+ 2 \cosh(2 |\Om| t))} }{\sqrt{\pi} \cosh(2)}
\left(  
\re^{ |\Om| t} \cosh \left( 2 \re^{- |\Om| t} {\cal X} \right) \sinh \left( 2 \re^{ |\Om| t} {\cal X} \right)
- \right .
\nonn
& ~~~~~~~~~~~~~~~~~~~~~~~~~~~~~~~~~~~~~~~~~~~~~~~~~~~~~~~~~~~~\left . 
\re^{- |\Om| t} \cosh \left( 2 \re^{ |\Om| t} {\cal X} \right) \sinh \left( 2 \re^{- |\Om| t} {\cal X} \right)
\right),
\nonn
& J_s({\cal X},t) = \frac{\hbar}{m} \frac{|\sigma|}{b_0}
\frac {\re^{- \uni (\th-\pi/4)} \re^{-({\cal X}^2+ 2 \cosh(2 |\Om| t))} }{\sqrt{\pi} \sinh(2)}
\left(  
\re^{ |\Om| t} \cosh \left( 2 \re^{|\Om| t} {\cal X} \right) \sinh \left( 2 \re^{ -|\Om| t} {\cal X} \right)
- \right.
\nonn
& ~~~~~~~~~~~~~~~~~~~~~~~~~~~~~~~~~~~~~~~~~~~~~~~~~~~~~~~~~~~~\left . 
\re^{- |\Om| t} \cosh \left( 2 \re^{ -|\Om| t} {\cal X} \right) \sinh \left( 2 \re^{|\Om| t} {\cal X} \right)
\right),
\nonn
& J_g({\cal X},t) = \frac{\hbar}{m} \frac{|\sigma|}{b_0}
\frac {\re^{- \uni (\th-\pi/4)} \re^{-({\cal X}- \cosh(|\Om| t))^2} }{\sqrt{\pi} }
2 \sinh(|\Om| t).
\end{flalign}
It is straightforward to prove that

\beqn
\partial_t (\rho_c(\th,x,t))+ \partial_x J_c(\th,x,t)& = & 0, \nonn
\partial_t (\rho_s(\th,x,t))+ \partial_x J_s(\th,x,t)& = & 0, \nonn
\partial_t (\rho_g(\th,x,t))+ \partial_x J_g(\th,x,t)& = & 0.
\eeqn

\subsection{Comparison of CSM with the use of the Gel'fand Triplet.} 

In \cite{nosswrhs}, we have studied the time evolution of an initial given state under the action of the Swanson Hamiltonian. We have shown that, for the region of the present study, one needs a Haussdorff vector space with a complex topology  $\Phi$, a Hilbert space $\cal H$ and the anti dual space of $\Phi$, namely $\Phi^\times$, to construct the Gel'fand triplet

\beqn
\Phi \in {\cal{H}} \in \Phi^\times.
\eeqn

Let us name $H^\times$ and $H_c^\times$ as the extensions of $H$ and $H_c$ on $\Phi^\times$, respectively. 
The generalised eigenfunctions of $H^\times$, $\tilde{g}^\pm_n(x)$, and of $H_c^\times$, $\ovl{g}^\pm_n(x)$, read\cite{nosswrhs}

\begin{equation}
\begin{split}
\tilde{g}^\pm_n(x)&={\rm e^{\frac {\alpha-\beta}{\omega-\alpha-\beta} \frac {x^2} {2 b_0^2}}}
g^\pm_n(x),~~~\wid{E}_n^\pm=E_n^\pm,\\
\bar{g}^\pm_n(x)&={\rm e^{-\frac {\alpha-\beta}{\omega-\alpha-\beta} \frac {x^2} {2 b_0^2}}}
\left(g^\pm_n(x)\right)^*, ~~~\ovl{E}_n^\pm=E_n^{\pm *},
\end{split}
\end{equation}
with $E_n^\pm =\hbar |\Omega| e^{\pm \textbf{i} \frac \pi 2} \left(n+\frac{1}{2}\right)$, and 
where $g^\pm_n(x)$ can be expressed as
\begin{equation}
g^\pm_n(x)= {\cal{N}}_n^\pm
\re^{-\re^{ \pm\textbf{i}\frac \pi 2}\frac{x^2}{2b_0^2}\vert \sigma\vert^2}H_n\left(\re^{\textbf{i}(\pm \frac \pi 4)}\frac{x}{b_0} |\sigma|\right), ~~~
{\cal{N}}_n^\pm=\sqrt{\dfrac{e^{\pm \textbf{i}\frac \pi 4}}{\sqrt{\pi} n! 2^n}\dfrac{|\sigma|}{b_0}}.
\end{equation}

As shown in \cite{nosswrhs}, the biorthogonal and the completeness relations read:

\begin{flalign}
&\left( {\ovl g}^\pm_m(x),\wid g^\pm_n(x) \right ) =
\int_C~({\ovl g}^\pm_m(x))^*\wid g^\pm_n(x) ~{\rm d}x =
\delta_{m n}, 
\label{bidsw}
\end{flalign}
and
\begin{flalign}
&\sum_{\begin{array}{c}
n=0 \\ s=\pm \\
\end{array}}^{\infty}~({\ovl g}^s_n(x))^*\wid g^s_n(x') = \delta(x-x'),
\label{compsw}
\end{flalign}
respectively.

The evolution operator acting on the generalised eigenfunctions of $\Phi^\times$ reads $U^\times(t)= \re^{\uni H^\times t}$\cite{maru}. For more details, the reader is kindly referred to \cite{nosswrhs}. Consequently, given the initial state $| I  \rangle$, its time evolution under $H^\times$, $|\wid I \rangle$, and under $H_c^\times$, $|\ovl I \rangle$, can be expressed as 

\begin{flalign}
&|\wid I (t) \rangle = |\wid I^+ (t)\rangle + |\wid I^- (t) \rangle, 
&|\ovl I (t) \rangle = |\ovl I^+ (t)\rangle + |\ovl I^- (t) \rangle,
\nonn
&|\wid I^+ (t) \rangle = \sum_n ~\re^{-\hbar |\Om| \left( n+\frac 12 \right)t } ~|\wid g_n^+ \rangle  \langle \ovl g_n^+  |I \rangle, 
&|\ovl I^+ (t) \rangle = \sum_n ~\re^{~\hbar |\Om| \left( n+\frac 12 \right)t } ~|\ovl g_n^+ \rangle  \langle \wid g_n^+  |I \rangle, 
\nonn
&|\wid I^- (t) \rangle = \sum_n ~\re^{~\hbar |\Om| \left( n+\frac 12 \right)t } ~|\wid g_n^- \rangle  \langle \ovl g_n^-  |I \rangle,
&|\ovl I^- (t) \rangle = \sum_n ~\re^{-\hbar |\Om| \left( n+\frac 12 \right)t } ~|\ovl g_n^- \rangle  \langle \wid g_n^-  |I \rangle. 
\nonn
\end{flalign}
In the coordinate representation, this split corresponds to the extension of the contour of the involved integrals to the upper and lower planes, respectively \cite{maru}.  

As an example, let us take a package of the form

\beqn
f(z)&=&\frac {\re^{-\frac{(z-2)^2}{2}}}{\pi^{1/4}},~~~z= \re^{\uni \frac \pi 4} |\sg| \frac  x {b_0},
\eeqn
so that 

\beqn
\wid f^+(x,0) & = & \sum_{n=0}^\infty \widetilde c^+_n 
~\wid g^+_n (x), ~~~{\wid c}^+_n = \frac{1}{n! \re \pi^{1/4} {\cal N}_n^+}, \nonn
\ovl f^+(x,0) & = & \sum_{n=0}^\infty \ovl c^+_n 
~\ovl g^+_n (x), ~~~{\ovl c}^+_n = ({\wid c}^+_n)^*, \nonn
\eeqn
Its time evolution for $t>0$ can be given in terms of the generalised eigenfunctions as

\beqn
\wid f(x,t) & = & \sum_{n=0}^\infty \widetilde c^+_n 
~\re^{-|\Om| (n+\frac 12 )t}~\wid g^+_n (x), \nonn
\ovl f(x,t) & = & \sum_{n=0}^\infty \ovl c^+_n 
~\re^{~|\Om| (n+\frac 12 ) t}~\ovl g^+_n (x). 
\eeqn
We can compute the survival probability, the probability of persistence, and the transmission and reflection coefficients by using the bi-orthogonal relation given in Eq. (\ref{bidsw}). In \ref{ap2}, we present the details of the computation. The results can be written as 

\beqn
{\cal P}^{-}(t) & = &  \bigg | \re^ { 2( \re^{- |\Om| t}-1)}  \bigg |^2,\\
Q(L,t) & =  & \frac 1 2 \left({\rm{Erf}} \left(L+2 \cosh(\Omt)\right)+{\rm{Erf}} \left(L-2 \cosh(\Omt)\right)\right).
\eeqn

The same result has been obtained in the framework of the CSM by taking $\theta=\pi/4$. 

\section{Conclusions}\label{conclusions}

In this work, we study the non-PT symmetry phase of a system modelled by the Swanson Hamiltonian \cite{swanson}. As shown in \cite{nosswrhs}, in the non-{\cal{PT}} symmetry phase the generalised eigenfunctions belong to the Rigged Hilbert Space, and are related to those of an inverted harmonic oscillator. By adopting the CSM, we perform a similar transformation, $V(\th)$, on the Hamiltonian $H$ of Eq.(\ref{hxp}). The eigenfunctions of the transformed Hamiltonian, $H(\th)$ of Eq.(\ref{htheta}), and of its adjoint operator, $H_c(\th)$ of Eq. (\ref{hthc}), are square-integrable functions in some regions of the model parameter space $(\th,\om,\al,\be)$. We construct the bi-orthogonal relations between the eigenfunctions of $H(\th)$ and $H_c(\th)$ \cite{noscsm}. To analyse the time evolution of a given initial wave package, we compute the response function associated with $H(\th)$ and $H_c(\th)$. As a first step, we compute the time evolution of the eigenfunctions of $H(\th)$ and $H_c(\th)$. The time evolution of a given initial package can be written in terms of the time-dependent eigenfunctions of the Hamiltonian. We define a Wigner Function according to the inner product proposed. We calculate the Wigner Functions associated with the eigenfunctions of $H(\th)$, both between eigenfunctions of the same eigenvalue and eigenfunctions with different eigenvalues. We derive the spatial density distribution, density current, and continuity equation from the Wigner Functions. Also, we compute the probability of persistence in the parabolic barrier and the Survival Probability of the initial wave packages. We discuss the results obtained for three particular wave packages and observe that they evolve in time as travelling waves.  As a final step, we compare the results obtained by adopting the CSM with the ones obtained by working in the Rigged Hilbert Space. As expected, for a particular value of the scaling parameter of the CSM, $\th= \pi/4$, both formalisms provide the same results. The advantage of using the CSM relies on working with square-integrable functions.

We have used the Swanson model as an example of the applicability of the CSM in the analysis of non-hermitian dynamics, particularly in describing the physics of EPs.  

Work is in progress concerning the evaluation of the Scattering Matrix using CSM for Swanson-like models.

\section{Acknowledgements}

This work was partially supported by the National Research Council of Argentine (PIP2023-2025, CONICET) and by the University of La Plata (11X/982-UNLP).


\appendix
\section{}\label{ap1}

The time evolution of the eigenfunctions of $H(\th)$ and $H(\th)_c$ can be found from the time evolution of the eigenfunctions of $\mathfrak{h}(\th)$ and $\mathfrak{h}(\th)_c$:

\beqn
\mathfrak{h}(\th) \phi^\pm_n(\th,x) & = & \pm \uni \hbar |\Om| \left (n+\frac 12  \right) \phi^\pm_n(\th,x), \nonn
\eeqn
and, if we define $\mathfrak{h}_c(\th)=\mathfrak{h}(-\th)=\mathfrak{h}(\th)^*$:
\beqn
\mathfrak{h}(-\th) \phi^\pm_n(-\th,x) & = & \pm \uni \hbar |\Om| \left (n+\frac 12  \right) \phi^\pm_n(-\th,x), \\
\label{h1}
\mathfrak{h}(\th)^* (\phi^\pm_n(\th,x))^* & = & \mp \uni \hbar |\Om| \left (n+\frac 12  \right) (\phi^\pm_n(\th,x))^*,
\label{h2}
\eeqn
from \eqref{h1} and \eqref{h2}, it results:

\beqn 
\phi_{c,n}^\pm(\th,x)=\phi^\pm_n(-\th,x)=(\phi^\mp_n(\th,x))^*, \nonn
E_{c,n}^\pm=\pm \uni \hbar |\Om| \left (n+\frac 12  \right).
\eeqn
and the basis bi-orthogonal is formed by noticing that:
\begin{flalign}
&  (\phi^\mp_{c,n}(\th,x), \phi^\pm_m(\th,x))=(\phi^\mp_n(-\th,x), \phi^\pm_m(\th,x))=((\phi^\pm_n(\th,x))^*, \phi^\pm_m(\th,x)) =\delta_{nm},\nonn
& ~~~~~~~~~~~~~~~~~~~~~~~~~~~~~~~~~~~~~~~~~~~~~~~~~~~~~~~~~~~
\int_{-\infty}^{\infty} \phi^\pm_n(\th,x) \phi_m^\pm(\th,x) {\rm{dx}} =\delta_{nm}.\nonn
\end{flalign}

Consequently, the time evolution of the eigenstates of $\mathfrak{h}(\th)$ and $\mathfrak{h}(\th)_c$ can be obtained as:
\beqn
\phi_n^\pm(\th,x,t)={\cal N}^\pm_n \int_{-\infty}^{\infty} K(\th,x,x',t) \phi_n^\pm(\th,x') {\rm{dx'}},
\eeqn
and 

\beqn
\phi^\mp_{c,n}(\th,x,t)=\phi^\mp(-\th,x,t)={\cal N}^\pm_n \int_{-\infty}^{\infty} K(-\th,x,x',t) (\phi_n^\pm(\th,x'))^* {\rm{dx'}}. \nonn
\eeqn
The normalization of the eigenstates is obtained from:

\begin{flalign}
({\cal N}^\pm_n)^2 \int_{-\infty}^{\infty} \int_{-\infty}^{\infty} \int_{-\infty}^{\infty} K(-\th,x,x',t)^* \phi^\pm_n( \th,x') K(\th,x,x'',t) \phi^\pm_n( \th,x'') {\rm {dx'}}{\rm {dx''}} {\rm {dx}} =1. \nonn
\end{flalign}

After some algebra, the time evolution of the eigenstates of $\mathfrak(\th)$ are given by:

\begin{flalign}
& \phi_n^-(\th,x,t)= \uni \re^{ -\uni \frac 12 n\pi} \re^{ -n  |\Om| t} \phi_n^-(\th,x),\nonn
& \phi_n^+(\th,x,t)= -\re^{ -\uni \frac 12 n \pi} \re^{ n  |\Om| t} \phi_n^+(\th,x). 
\end{flalign}
and those of $\mathfrak(-\th)$ read:
\begin{flalign}
& \phi_{c,n}^+(\th,x,t)= \re^{ \uni n\pi} \phi_n^+(-\th,x,t), \nonn
& \phi_{c,n}^-(\th,x,t)= \re^{ \uni n\pi} \phi_n^-(-\th,x,t).
\end{flalign}

In terms of the eigenfunctions of $H(\th)$, the components of the Wigner function can be computed as

\beqn 
W_{m n}^\pm(\th,x,p)=\int_{-\infty}^{\infty}~\ovl{\psi}_m^\pm \left( \th,x+\frac y 2,t \right)^*~\wid{\phi}_n^\pm \left(\th,x-\frac y 2,t\right)~\re^{\uni \frac{p} {\hbar }y }~{\rm {dy}}.
\eeqn
After some algebra, it reads 
\beqn 
W_{m n}^-(u,v) & = & 
2 \sqrt{\frac{m!}{n!}} \re^{- (n-m)|\Om|t}\re^{-u v} u^{n-m}{\rm L }_m^{n-m}(~2 u v), ~~~ m<n,\nonn
& = &
2 \sqrt{\frac{n!}{m!}} \re^{- (n-m)|\Om|t}\re^{-u v} v^{m-n}{\rm L }_n^{m-n}(~2 u v), ~~~ n<m ,\nonn
& = &
2 \re^{-u v} {\rm L }_n(~2 u v), ~~~ n=m ,\nonn
W_{m n}^+(u,v) & = & 2 \sqrt{\frac{m!}{n!}} \re^{~(n-m)|\Om|t}\re^{~ u v} v^{n-m}{\rm L }_m^{n-m}(-2 u v), ~~~ m<n,\nonn
& = & 2 \sqrt{\frac{n!}{m!}} \re^{~(n-m)|\Om|t}\re^{~ u v} u^{m-n}{\rm L }_n^{m-n}(-2 u v),~~~ n<m ,\nonn
& = &
2 \re^{~u v} {\rm L }_n(-2 u v), ~~~ n=m ,\nonn
u (\th,x,p) & = &  \lpth-\frac {\alpha-\beta}{(\omega-\alpha-\beta)|\sg|^2} \lth -\uni \lth, \nonn
v (\th,x,p) & = &  \lpth-\frac {\alpha-\beta}{(\omega-\alpha-\beta)|\sg|^2} \lth +\uni \lth. \nonn
\eeqn

\section{}\label{ap2}

The density distribution of Eq. (\ref{pack}), can be written also as

\beqn
\rho(\th,x,t)= \ovl{f}^*(\th,x,t) \wid{f}(\th,x,t), 
\eeqn
so that 

\beqn
\partial_t(\rho(\th,x,t))= \partial_t(\ovl{f}^*(\th,x,t)) \wid{f}(\th,x,t)+\ovl{f}^*(\th,x,t) \partial_t(\wid{f}(\th,x,t)) , 
\eeqn
if we use the fact that:

\beqn
 \uni \hbar~ \partial_t (\wid{f}(\th,x,t)) & = & {\rm H}(\th) \wid{f}(\th,x,t), \nonn
 \uni \hbar~ \partial_t (\ovl{f}(\th,x,t)) & = & {\rm H}_c(\th) \ovl{f}(\th,x,t), \nonn
-\uni \hbar~ \partial_t (\ovl{f}^*(\th,x,t)) & = & {\rm H}_c^*(\th) \ovl{f}^*(\th,x,t) \nonn
& = & {\rm H}(\th) \ovl{f}^*(\th,x,t).
\eeqn
then

\beqn
\partial_t(\rho(\th,x,t))&=& -\frac{1}{\uni \hbar} ~{\rm H}(\th)( \ovl{f}^*(\th,x,t)) \wid{f}(\th,x,t)+
\frac{1}{\uni \hbar} ~\ovl{f}^*(\th,x,t) {\rm H}(\th)( \wid{f}(\th,x,t)), \nonn
&=& \frac{\hbar}{2 \uni m} ~ \left( \frac{{\partial}^2 \ovl{f}^*}{\partial{\rm {x}}^2}\wid{f}-
 ~\ovl{f}^*\frac{{\partial}^2 \wid{f}}{\partial{\rm {x}}^2} \right), \nonn
 &=& \frac {\partial}{\partial{\rm x}} \left( \frac{\hbar}{2 \uni m} ~ \left( \frac{{\partial} \ovl{f}^*}{\partial{\rm {x}}}\wid{f}-
 ~\ovl{f}^*\frac{{\partial} \wid{f}}{\partial{\rm {x}}} \right)\right), \nonn
 &=& -\frac {\partial}{\partial{\rm x}} \left( J (\th,x,t)\right), \nonn
\eeqn
thus

\beqn
\frac {\partial}{\partial{\rm t}}\left(\rho(\th,x,t)\right)+\frac {\partial}{\partial{\rm x}} \left( J (\th,x,t)\right)=0.
\eeqn

\end{document}